\documentclass[aps,prc,twocolumn,tightenlines,ece,amsmath,groupedaddress,showpacs,showkeys]{revtex4}
\usepackage{epsf}
\usepackage{graphicx}
\usepackage{graphics}
\usepackage{color}
\def\beg{\begin{equation}}
\def\ee{\end{equation}}

\usepackage{longtable}
\begin{document}
\def\prl{{\em Phys. Rev. Lett. }}
\def\prc{{\em Phys. Rev. C }}
\def\jap{{\em J. Appl. Phys. }}
\def\ajp{{\em Am. J. Phys. }}
\def\nima{{\em Nucl. Instr. and Meth. Phys. A }}
\def\npa{{\em Nucl. Phys. A }}
\def\npb{{\em Nucl. Phys. B }}
\def\epjc{{\em Eur. Phys. J. C }}
\def\plb{{\em Phys. Lett. B }}
\def\mpla{{\em Mod. Phys. Lett. A }}
\def\pr{{\em Phys. Rep. }}
\def\prv{{\em Phys. Rev. }}
\def\zpc{{\em Z. Phys. C }}
\def\zpa{{\em Z. Phys. A }}
\def\ppnp{{\em Prog. Part. Nucl. Phys. }}
\def\jpg{{\em J. Phys. G }}
\def\cpc{{\em Comput. Phys. Commun.}}
\def\app{{\em Acta Physica Pol. B }}
\def\aip{{\em AIP Conf. Proc. }}
\def\jhep{{\em J. High Energy Phy. }}
\def\ijmpa{{\em Int. J. Mod. Phys. A }}

% Use the \preprint command to place your local institutional report
% number in the upper righthand corner of the title page in preprint mode.
% Multiple \preprint commands are allowed.
% Use the 'preprintnumbers' class option to override journal defaults
% to display numbers if necessary
%\preprint{}
%\markboth{{Constituent Quarks and Multi-strange Baryon Production}}{{ Nirbhay K. Behera, Raghunath Sahoo and B.K. Nandi}}
%Title of paper
%\title{Constituent Quarks and Enhancement of Multi-strange Baryons in Heavy-Ion Collisions}
\title{Constituent Quark Scaling of Strangeness Enhancement in Heavy-Ion Collisions}
% repeat the \author .. \affiliation  etc. as needed
% \email, \thanks, \homepage, \altaffiliation all apply to the current
% author. Explanatory text should go in the []'s, actual e-mail
% address or url should go in the {}'s for \email and \homepage.
% Please use the appropriate macro foreach each type of information

% \affiliation command applies to all authors since the last
% \affiliation command. The \affiliation command should follow the
% other information
% \affiliation can be followed by \email, \homepage, \thanks as well.

\author{Nirbhay K. Behera{$^1$}, Raghunath
  Sahoo{$^2$}\footnote{Corresponding Author, Email:
    Raghunath.Sahoo@cern.ch}, 
Basanta K. Nandi{$^1$}} 

%\email[]{Raghunath.Sahoo@pd.infn.it}

%\homepage[]{Your web page}
%\thanks{}
%\altaffiliation{}
%\affiliation{{$^1$}Dipartimento di Fisica dell'Universit$\grave{a}$
%and Sezione INFN di Padova, Italy}
\affiliation{{$^1$}Indian Institute of Technology Bombay, Mumbai, India-400067}
\affiliation{{$^2$}Indian Institute of Technology Indore, Indore, India-452017}

%Collaboration name if desired (requires use of superscriptaddress
%option in \documentclass). \noaffiliation is required (may also be
%used with the \author command).
%\collaboration can be followed by \email, \homepage, \thanks as well.
%\collaboration{}
%\noaffiliation

\date{\today}

\begin{abstract}
In the frame work of a nuclear overlap model, we estimate the number of 
nucleon and quark participants in proton-proton, proton-nucleus and 
nucleus-nucleus collisions. We observe the number of
nucleon ($N_{N-part}$)-normalized enhancement of multi-strange particles which 
show a monotonic increase with centrality, turns out to be a centrality independent scaling behavior
when normalized to number of constituent quarks participating in the 
collision ($N_{q-part}$). In addition, we observe that the $N_{q-part}$-normalized enhancement, 
when further normalized to the strangeness content, shows a strangeness 
independent scaling behavior. This holds good at top RHIC energy. However, the corresponding SPS 
data show a weak $N_{q-part}$-scaling with strangeness scaling being violated at top SPS energy.
This scaling at RHIC indicates that the partonic degrees of 
freedom playing an important role in the production of multi-strange 
particles. Top SPS energy, in view of the above observations, shows a co-existence of hadronic and partonic phases.  We give a comparison of data with HIJING, AMPT and UrQMD models to understand the particle production dynamics at different energies. 

\end{abstract}

% insert suggested PACS numbers in braces on next line
% insert suggested keywords - APS authors don't need to do this

%\maketitle must follow title, authors, abstract, \pacs, and \keywords
\maketitle

% body of paper here - Use proper section commands
% References should be done using the \cite, \ref, and \label commands
%\section{}
% Put \label in argument of \section for cross-referencing
%\section{\label{}}
%\subsection{}

\maketitle
\section{INTRODUCTION} 
Relativistic heavy-ion collisions aim at creating matter at extreme conditions of 
energy density and temperature which is governed by the partonic degrees
of freedom called Quark-Gluon Plasma (QGP). The main focus of these studies is
the observation of quark-hadron phase transition and exploring the
Quantum Chromodynamics (QCD) phase 
diagram. In the early phases of ultra-relativistic heavy-ion collisions, when
a hot and dense region is formed in the core of the reaction zone, different
quark flavors are produced. Then the produced matter undergoes transverse
expansion and multiple scattering among the produced  particles.
The formation of the hadrons from the partonic phase is accomplished through
further expansion and cooling of the system. In proton-proton ($p+p$) collisions, the formation of
QGP is not expected, whereas a possible formation of  QGP is
expected in nucleus-nucleus (A+A) collisions. Hence, a comparative
study of produced particles in A+A collisions, with that of $p+p$
collisions, could give better understanding of the properties of the medium formed in A+A
collisions.

In the mid-rapidity region, strangeness enhancement has been proposed as a 
potential signature of QGP \cite{pr,larry}. Strange baryons are produced in 
strong interaction processes and decay through weak interaction. It has 
been observed that multi-strange baryons, {\it i.e.} $\Omega (sss)$, $\Xi(ssd)$ and 
$\Lambda (uds)$,  are formed and decouple from the 
system earlier in time \cite{nu}. Due to their different reaction rates in the
medium, particles with different strangeness decouple at
different times. Relativistic Quantum Molecular Dynamics (RQMD) results suggest
that the multi-strange baryons freeze-out at energy densities more than 
$1 ~GeV/fm^3$ \cite{nu} which corresponds to the critical energy density 
predicted by lattice QCD calculations \cite{lattice}.
The study of multi-strange baryons are of paramount importance in high energy 
heavy-ion collisions because of their dominant strangeness content (s-quark). 
As the colliding species in nuclear collisions do not contain any strange valence quark, 
particles with non-zero strange quarks can only be produced out of the collision process.
The production of strange particles is enhanced in a QGP phase compared to 
a hadronic system. This is because, the production rate of $gg \rightarrow s\bar{s}$ (gluon fusion)
is high in a QGP medium \cite{gluglu}, which is absent in hadronic phase. 
In addition, multi-strange baryons are less suffered by hadronic rescatterings 
in the later stage of the
evolution of the fireball because of their small hadronic interaction
cross sections. That is why multi-strange hadrons are good probes to carry the early stage 
information \cite{van,biagi,cheng,bass,mullerRA}. Hydrodynamic model
estimations on hadron $p_T$ spectra suggest that the thermal
freeze-out temperature of multi-strange baryons are close to their chemical freeze-out
temperature ($T_{ch} \sim 160 $ MeV), which is around the critical temperature, $T_c$,
for deconfinement transition. This indicates that multi-strange baryons are almost not
affected by hadronic rescatterings at the later stage of the heavy-ion collisions \cite{van,bass,bassDumitru}.
Hence, multi-strange baryons could carry the information of  possible formation of  a QGP phase.
It could be envisaged that the multi-strange baryons are formed 
out of partonic interactions rather than nucleonic interactions. We shall be justifying this in 
the following sections.

In this paper, both nuclei and nucleons are considered as superposition of
constituent or ``dressed'' quarks (partons or valons). Baryons are composed of 
three and mesons are of two such quarks.
 The concept of constituent quarks is very well known \cite{hwa, aniso}
and established in the realm of the discovery of constituent quark scaling of 
identified particles elliptic flow at RHIC \cite{starFlow}. The constituent quark
approach is successful in explaining many features of hadron-hadron, hadron-nucleus
\cite{aniso1} and nucleus-nucleus collisions. These include global properties like 
the charged particle and transverse energy density per participant pair 
\cite{voloshin, bm}. QCD calculations support the presence of 
three objects of size 0.1-0.3 fm inside a nucleon \cite{shuryak}. Furthermore, 
it has been seen
that nucleus-nucleus collisions and $p+p$ collisions have similar initial states
if the results are scaled by the number of constituent quark participants 
\cite{rachid}. These observations also indicate that the particle production is
essentially controlled by number of constituent quarks pairs participating in 
the collision.
 In a constituent quark picture, nucleon-nucleon ($NN$) collision looks like 
a collision of two light
nuclei with essentially one $qq$ pair interacting in the collision, leaving other
quarks as spectators. These quark spectators form hadrons in the nucleon
fragmentation region with a part of the entire nucleon energy being used for 
the particle production ($\sqrt{s_{qq}}\sim \sqrt{s_{NN}}/3$). But in
A+A collisions, due to the large size of the nucleus compared to the
nucleon, there is a higher probability of $q-q$ interaction between
the projectile and target nucleons. It has been observed at RHIC energies 
that the production rate of strange and multi-strange baryons in Au+Au collisions
at $\sqrt{s_{NN}}= 200$ GeV, when scaled by the number of nucleon participants,
 are different (get enhanced) when compared to similar measurements in $p+p$ 
collisions at  same energy \cite{strangeStar}. The observed enhancement increases 
 with strangeness content of the baryons and also as a function of collision
 centrality, while going from peripheral to central collisions. Similar observations
 have been made at SPS energy for Pb+Pb collisions at $\sqrt{s_{NN}}= 17.3$ GeV,
 when compared with the corresponding measurements for p+Be collisions at the 
 same energy \cite{spsStrange}. In this paper, we
have compared the centrality dependence of number of nucleon
participant normalized multi-strange baryon
enhancement at SPS and RHIC energies with the expectations from
HIJING, AMPT and UrQMD models. The linear rise of the enhancement seen
in the data, gets converted into a spectacular quark-participant
scaling behavior when normalized to number of
quark-participants. Furthermore, we explore the strangeness scaling,
where quark-participant normalized enhancement for different
multi-strange baryons, when divided by the strangeness content show a
strangeness-independent scaling behavior. 

\section{CALCULATION OF THE NUMBER OF PARTICIPANTS}
The calculations of the mean number of nucleon/quark participants are done in
the following way. In the nuclear overlap model, the mean number of participants, {\it i.e.} $N_{N-part}$, in 
the collisions of a nucleus $A$ and a nucleus $B$ with impact parameter $b$ is given by
\begin{align}
N_{N-part, AB} = \int d^2s~T_A(\vec{s})\lbrace 1-\lbrack 1-\frac{\sigma_{NN}T_B(\vec{s}-
\vec{b})}{B}\rbrack ^B\rbrace \nonumber\\
+ \int d^2s~T_B(\vec{s})\{1-[1-\frac{\sigma_{NN}T_A(\vec{s}-\vec{b})}{A}]^A\},
\label{npart}
\end{align}
{where $T(b)= \int_{-\infty}^{\infty} dz ~n_A(\sqrt{b^2+z^2})$ is the thickness function,
defined as the probability of having a nucleon-nucleon ($NN$) collision 
within the transverse area element $db$. 
$[1-\sigma_{NN}T_A(b)/A]^A$ is the probability for a nucleon to 
pass through the nucleus without any collision. $A$ and $B$ are the mass numbers of
two nuclei participating in the collision process. We use the Woods-Saxon nuclear density profile,
\begin{equation}
n_A(r) = \frac{n_0}{1+exp[(r-R)/d]},
\label{density}
\end{equation}
with parameters, the normal nuclear density $n_0 = 0.17 ~ fm^{-3}$, the nuclear
radius $R = (1.12 A^{1/3} - 0.86^{-1/3})~fm$  and the skin depth $d = 0.54~fm$.
The inelastic nucleon-nucleon cross sections, {\it i.e.} $\sigma_{NN}$,
are 42 ~mb at $\sqrt{s_{NN}} = 200 ~ GeV$ and 30 ~mb at $\sqrt{s_{NN}} = 17.3 ~ GeV$.} 
Considering proton as a point particle and nucleus as an extended
object in a $p+A$ collision, the number of participating nucleons is
given by 
\begin{align}
N_{N-part, pA} = \lbrace 1-\lbrack 1-\frac{\sigma_{NN}T_A(b)}{A}\rbrack
^A\rbrace + T_A(b) \sigma_{NN}.
\label{npart-pA}
\end{align}

In order to calculate the number of quark participants, $N_{q-part}$, in nucleus-nucleus
collisions, the density for
quarks inside the nucleus is changed to three times that of the nucleon density
($n_0^q = 3 n_0= 0.51 ~ fm^{-3}$). Instead of nucleon-nucleon cross section, quark-quark 
cross section is used which is $4.67~mb$ and $3.3~ mb$ at
$\sqrt{s_{NN}}=$ 200 ~GeV and 17.3 GeV, respectively \cite{voloshin, de}. 
In $p+p(\bar{p})$ collisions, the quark participants are calculated by considering the proton
and antiproton as hard sphere of radius 0.8 $fm$ \cite{wong}.
And for asymmetric collisions, like $p+A$ collisions,
proton is considered as a hard sphere of radius 0.8 $fm$ and nucleus as extended objects
with a Woods-Saxon density profile.

\section{RESULTS AND DISCUSSION}
At lower center of mass energies, it has been found that the particle production 
scales with the number of participating nucleons, contrary to the case of high 
energies where hard processes dominate. Hard processes have much smaller 
cross-section than the soft processes. However, the number of binary collisions 
increase with increase in collision centrality faster than the number of 
participants. As a result, the particle production per participant nucleon 
increases with centrality. By using constituent quark approach, 
we are going to show how the particle production at higher energies depends on 
the participating quarks. For this, multi-strange particles are chosen 
because of the proposed signature of strangeness enhancement in QGP medium. 
Few of the mechanisms for strangeness enhancement include:\\
~~~(i) The chemical and flavor equilibration time in gluon rich plasma have been
predicted to be shorter than a thermally equilibrated hadronic matter of $T \sim 160$ MeV \cite{gluglu}.  
Gluon fusion ($gg \rightarrow s\bar{s}$) is a dominant production mechanism of strangeness 
in an equilibrated gluon rich plasma \cite{gluglu}. This might allow for strangeness equilibration within the lifetime 
of QGP and hence resulting in strong enhancement of strangeness compared to a hadronic system.
This enhancement in strangeness can also be explained in the context of statistical mechanics. 
In $p+p$ system, the strangeness enhancement is not 
expected as the available volume is much smaller compared to nucleus-nucleus system. 
The $p+p$ system can be treated as  
canonical system and  the net strangeness number should be conserved. 
So the s$\bar{s}$ pairs are created at the same point and are
annihilated as well, resulting in the suppression of strange
hadrons. But in Au+Au system, strange and anti-strange hadrons  
are created independently and statistically distributed over the entire 
nuclear fireball, which could be treated in a grand canonical ensemble
approach (GC) \cite{becattini}.  As the system thermalizes, the phase space suppression disappears 
as the volume available is more. The volume here is linearly proportional to the 
number of participating nucleons, {\it i.e.} $N_{N-part}$. So a relative strangeness 
enhancement is observed in Au+Au collisions with respect to $p+p$ collisions \cite{strangeStar}. \\
 (ii) Early stage multiple scattering in heavy-ion collisions, may lead to an increase of the color 
electric field strength. 	In string-hadronic model, particles are produced through fragmenting color field (string) in Schwinger mechanism \cite{schwinger}. In high energy heavy-ion collisions, the string density could be very 
high so that color flux tubes overlap leading to a superposition of color electric fields. 
This results in enhanced production of multi-strange particles. This mechanism has been verified in the 
framework of UrQMD model with $\Omega$ showing an enhancement factor up to 100 \cite{bleicher}.\\

%Fig1
%\begin{figure}
 %\includegraphics[scale=0.4]{naprtVsCentrality-Fig1.eps}\\
%\includegraphics[scale=0.45]{Fig1.eps}\\
% \caption{\small (Color online) Mean number of nucleon participants as a function of centrality in 
 %  the overlap model (filled square) and from STAR calculations (filled circles).}
 %\label{npartCen}
%\end{figure}
%Fig2
%\begin{figure}
%\includegraphics[scale=0.65]{101008PRLLamdaFig.eps}\\[-0.25cm]
%\includegraphics[scale=0.4]{nNpartByQpartVsNnpart-Fig2.eps}\\
%\includegraphics[scale=0.4]{Fig2.eps}\\
%\caption{\small (Color online) The ratio of $N_{q-part}/N_{N-part}$ as a function of centrality for %Au+Au collisions at $\sqrt{s_{NN}} = 200$ GeV from the 
 % overlap model calculations.}
%\label{nQByNp}
%\end{figure}
%\begin{figure}
%Fig3
%\includegraphics[scale=0.4]{Fig3.eps}\\
%\caption{\small (Color online) Mean number of nucleon and quark participants as a function of impact %parameter
 %from the overlap model calculations for Pb+Pb  and Au+Au collisions at SPS and RHIC top 
 %energies.}
%\label{npartIP}
%\end{figure}
To study the constituent quarks dependence of strangeness enhancement, we need to 
estimate the number of participating quarks, which has been done in the framework 
of nuclear overlap model. It is very essential to check how good is our
estimation of number of participating nucleons in the collision.
In order to do that, the mean number of participating nucleons,
calculated in overlap model, are compared with the number estimated by
the STAR experiment \cite{starNpart}. A very good agreement of nuclear 
overlap model calculations with that of STAR estimations has been observed.
We have then estimated the number of quark participants within the
prescription described in the previous section. The ratio of quark
partipants and nucleon participants has been found to increase monotonically
with collision centrality \cite{voloshin}.  This ratio shows a sharp increase for $N_{N-part} \leq ~ 100$, which
follows a type of linear rise while going from peripheral to central collisions. 
%Fig.4
\begin{figure*}
\begin{center}
\includegraphics[width=86mm]{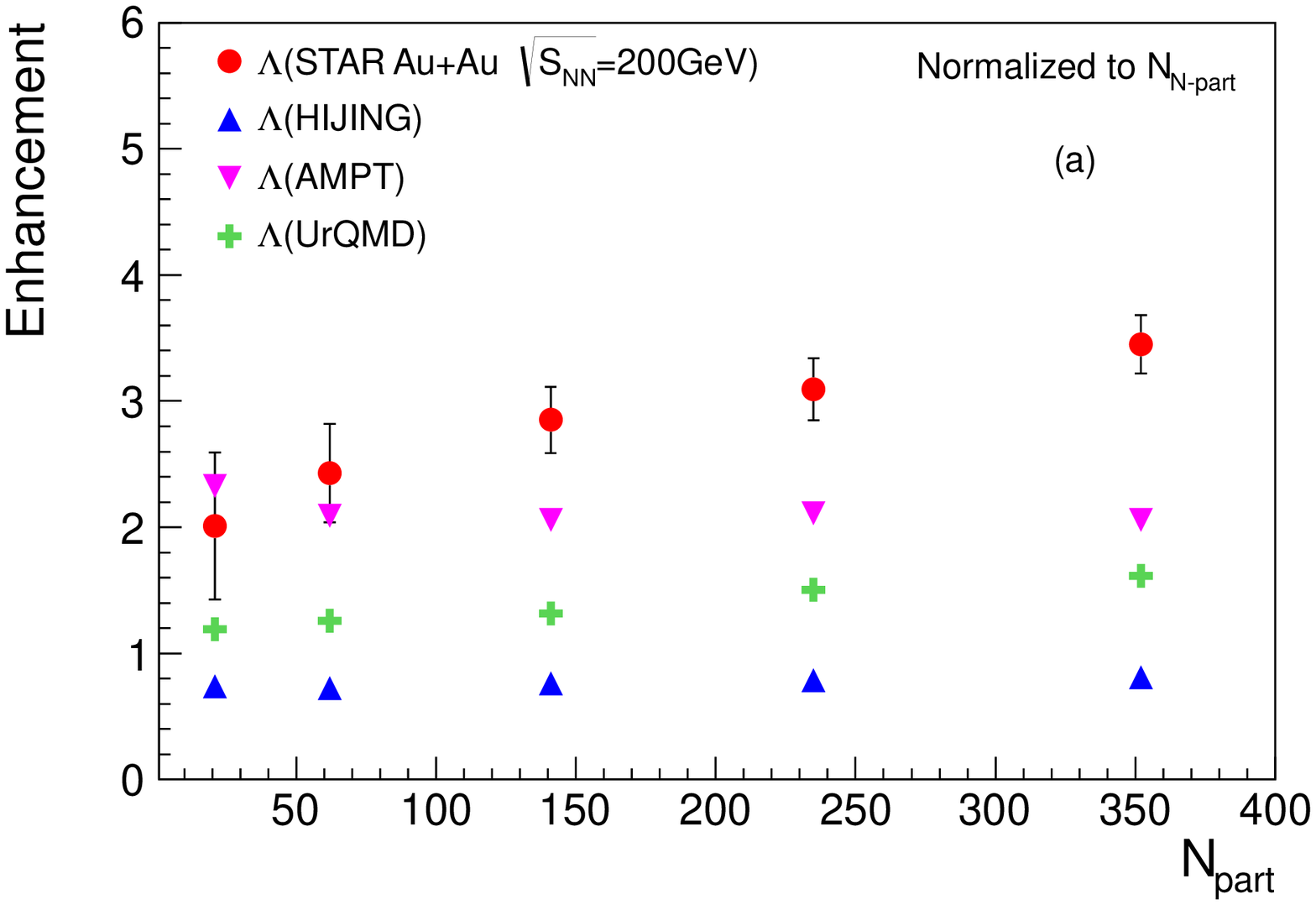}
\includegraphics[width=86mm]{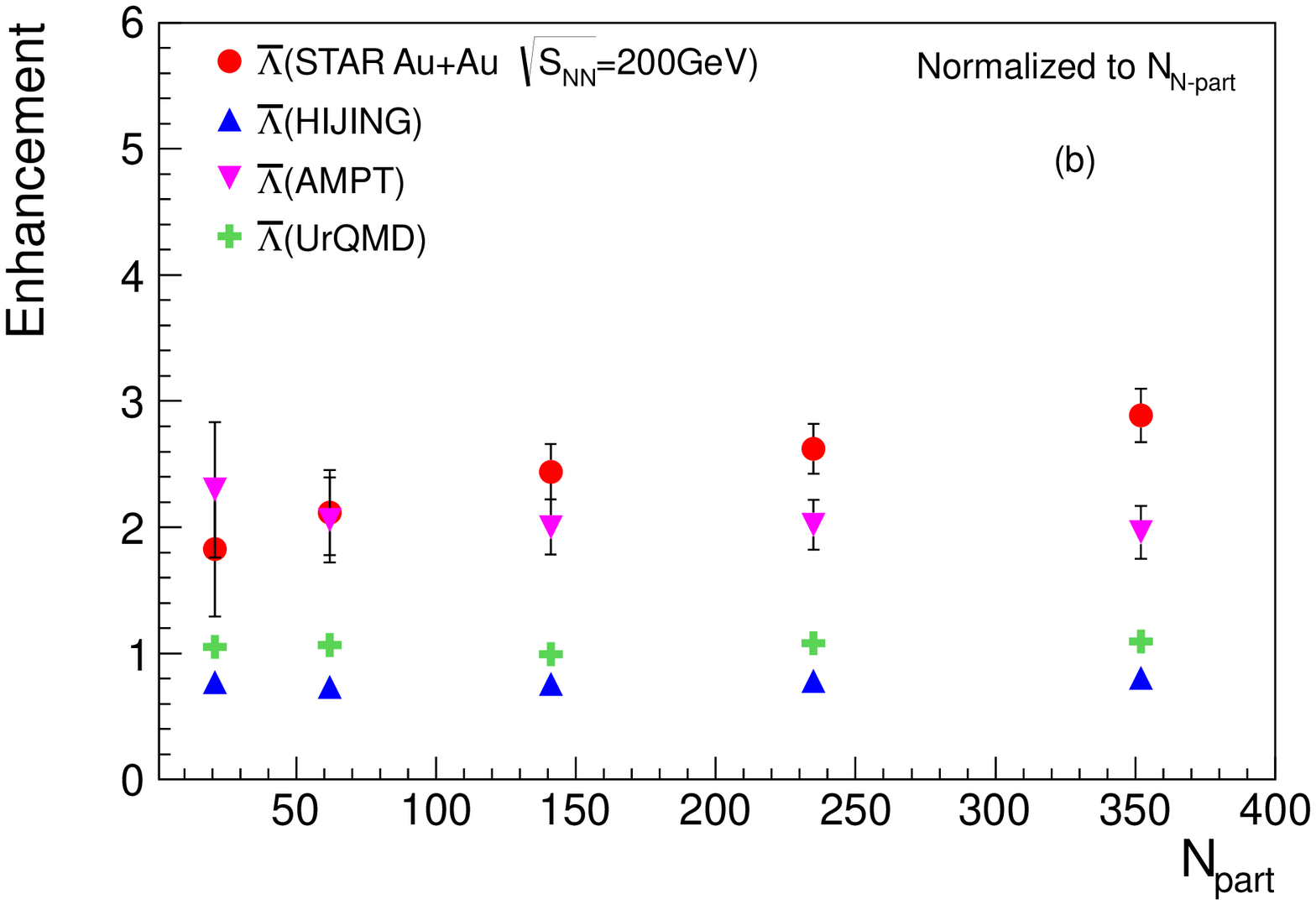}\\
\caption{\small (Color online) $N_{N-part}$-normalized
enhancement of (a) $\Lambda$ and (b) $\bar{\Lambda}$ as a
 function of collision centrality for Au+Au collisions at $\sqrt{s_{NN}} = 200$
 GeV are compared with HIJING, AMPT and UrQMD models at mid-rapidity.}
\label{lambdaSimRhic}
\end{center}
\end{figure*}

%Here,  an attempt has been made to explain the findings of STAR experiment on 
%the enhancement of multi-strange baryons as a function of collision centrality, 
%in heavy-ion collisions compared to $p+p$ collisions, given in Figure 1 
%of Ref. \cite{strangeStar}.  For each particle species $i$, the yield 
%enhancement factor, $E(i)$, is given by

The yield enhancement factor, $E(i)$, for particle species $i$ is given by
\begin{equation}
E(i) = \frac{Yield^{AA}(i) \langle N_{N-part}^{NN}
  \rangle}{Yield^{NN}(i) \langle N_{N-part}^{AA} \rangle},
\label{enhancement}
\end{equation}
where, $Yield^{AA}(i)$  and $Yield^{NN}(i)$ are the yields of strange particles, and  $N^{AA}_{N-part}$ and $N^{NN}_{N-part}$
are the number of nucleon participants, in nucleus-nucleus  and nucleon-nucleon collisions, respectively.
The number of nucleon participants, {\it i.e.} $N_{N-part}$, 
is used to characterize the collision centrality. However, in the constituent quark 
framework, the nucleon participants no longer bear the meaning of sources of particle 
production. It is assumed that in this picture the constituent quarks are participating 
in the reaction and are the sources of interest for particle production. To understand 
the collision
dynamics, the collision data are compared with models like HIJING-1.35, 
AMPT (default version) and UrQMD-3.3p1. 
HIJING is a Heavy Ion Jet Interaction Generator  which includes multiple minijet 
production, nuclear shadowing of parton distribution functions and the mechanism 
of jet interaction with dense matter. This model is based on perturbative 
QCD \cite{hijing,hijing1}. Motivated by perturbative QCD,  AMPT (A Multi-Phase Transport)
 model includes both initial partonic and final state hadronic interactions with 
quark-gluon to hadronic matter transition and their space-time evolution. Here,
the partons are allowed to undergo scattering before they hadronize \cite{ampt0,ampt}. 
UrQMD (Ultra-relativistic Quantum Molecular Dynamics) model  is based on microscopic 
transport theory, where hadronic interactions play an important role describing 
the evolution of the system \cite{urqmd}.

In Figure \ref{lambdaSimRhic}, $N_{N-part}$-normalized enhancement of
(a) $\Lambda$ and (b) $\bar{\Lambda}$-baryons for Au+Au collisions 
at $\sqrt{s_{NN}} = 200 $ GeV 
are shown as a function of $N_{N-part}$. This collision data are from the STAR 
experiment \cite{strangeStar} and are compared
with the corresponding estimates of HIJING, AMPT and UrQMD models. In this paper, the $p+p$ yield is taken from experimental data for the enhancement factor calculations.  It
is observed that none of these models correctly describes the collision data. All the models, {\it i.e.} 
HIJING, AMPT and UrQMD, show a centrality independent behavior. 
It is observed that the $N_{N-part}$-normalized enhancement, obtained from AMPT model,
is higher than the data obtained from HIJING and UrQMD models. This is because of the initial state partonic interactions in AMPT model.

%Fig5
\begin{figure*}
\begin{center}
\includegraphics[width=86mm]{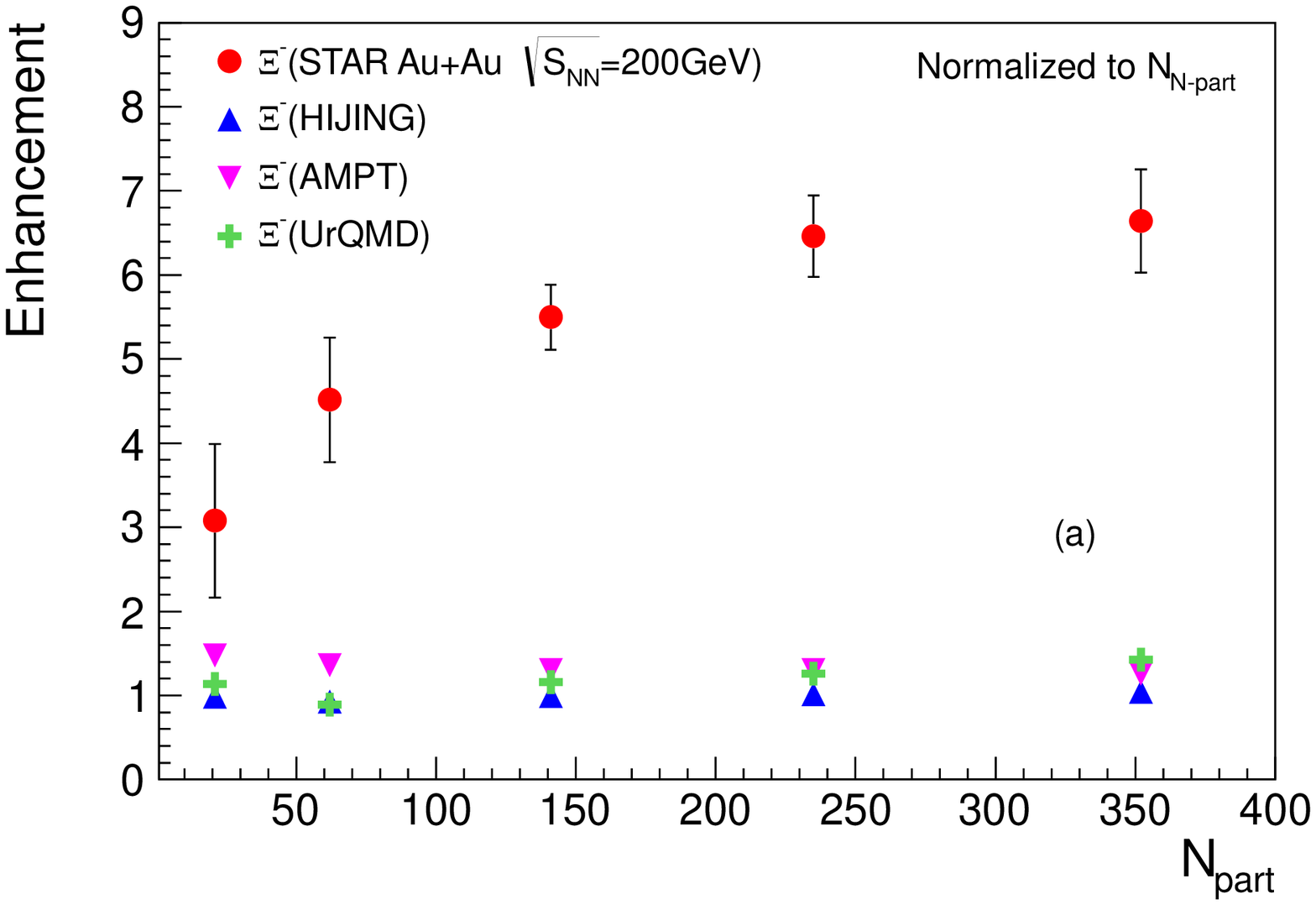}
\includegraphics[width=86mm]{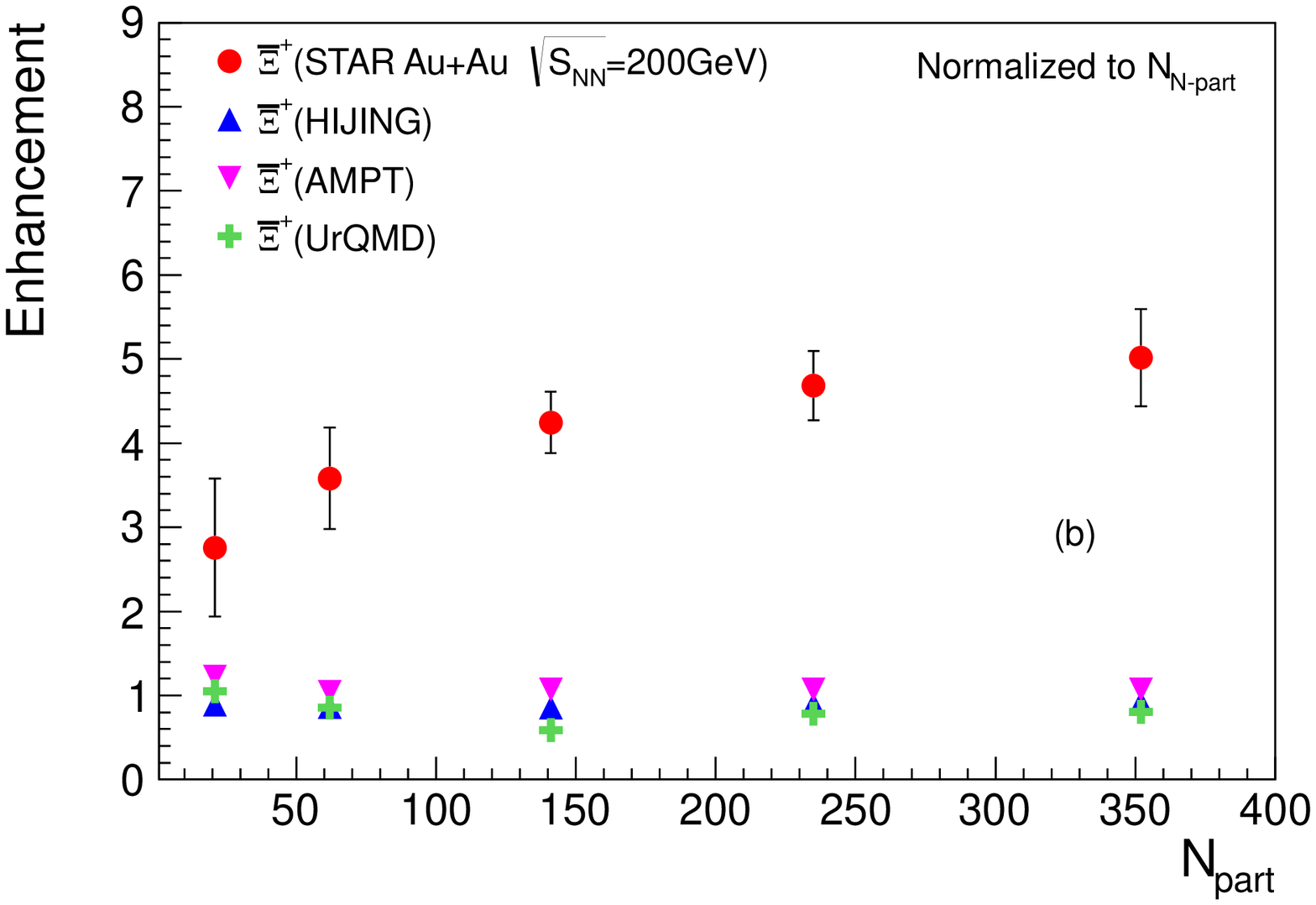}\\
\caption{\small (Color online) $N_{N-part}$-normalized
 enhancement of  (a) $\Xi^-$ and (b) $\bar{\Xi}^+$ as a function of
  collision centrality for Au+Au collisions at $\sqrt{s_{NN}} = 200$
  GeV are compared with HIJING, AMPT and UrQMD models at mid-rapidity.}
\label{xiSimRhic}
\end{center}
\end{figure*}

In Figure \ref{xiSimRhic}, $N_{N-part}$-normalized enhancement of
(a) $\Xi^-$ and (b) $\bar{\Xi}^+$-baryons at $\sqrt{s_{NN}} = 200 $ GeV for Au+Au collisions are
shown  as a function of $N_{N-part}$. The collision data points, obtained from STAR experiment
\cite{strangeStar}, are compared with HIJING, AMPT and UrQMD models.
The enhancement observed from the model estimates are independent of centrality
and are much less than the data (enhancement factor being close to one). Although AMPT 
model has partonic degrees of freedom, the enhancement of $\Xi^-$ and $\bar{\Xi}^+$ being 
close to one, like HIJING and UrQMD models,  indicate that AMPT model doesn't describe the 
enhancement of baryons of higher strangeness content.

\begin{figure}
%Fig6
\includegraphics[width=86mm]{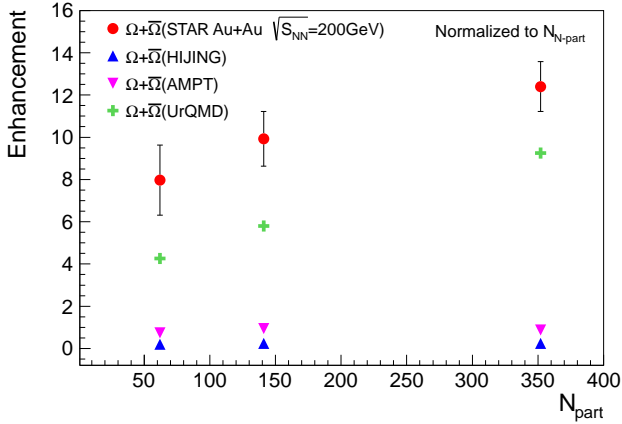}\\
\caption{\small (Color online) $N_{N-part}$-normalized
 enhancement of $\Omega+\bar{\Omega}$ as a function of collision
  centrality for Au+Au collisions at $\sqrt{s_{NN}} = 200$
  GeV are compared with HIJING, AMPT and UrQMD models at mid-rapidity.}
\label{omegaSimRhic}
\end{figure}

The collision data obtained from STAR experiment \cite{strangeStar} on 
$N_{N-part}$-normalized enhancement of
$\Omega +\bar{\Omega}$-baryons at $\sqrt{s_{NN}} = 200 $ GeV for Au+Au collisions 
are shown in Figure \ref{omegaSimRhic}  as a function of $N_{N-part}$ and are compared
with the corresponding estimates of HIJING, AMPT and UrQMD models. The enhancement 
observed from HIJING and AMPT models  are independent of centrality and the enhancement 
factor is  close to one.  Whereas the corresponding estimates from UrQMD show a rise 
with centrality like experimental data, only difference being the enhancement factor 
is higher in case of data compared to UrQMD.  A close observation of UrQMD 
estimation of $N_{N-part}$-normalized enhancement of strange-baryons at 
$\sqrt{s_{NN}} = 200 $ GeV for Au+Au collisions (shown in 
Figure \ref{lambdaSimRhic} - \ref{omegaSimRhic})
shows a strangeness dependent increase in the enhancement. This means that 
the $N_{N-part}$-normalized enhancement increases monotonically while going 
from $\Lambda$ to $\Omega$-baryons. This monotonic behavior of strangeness 
enhancement seems interesting in view of UrQMD being a microscopic trasport 
model with hadronic scattering playing a vital role in describing the space-time 
evolution of the system. Multi-strange baryons are produced early in time and also
their freeze-out time is very small \cite{nu}. With hydro+UrQMD calculations
and taking hadronic rescattering into account, the mean freeze-out times for $\Omega,
\Xi$ and $\Lambda$ at top RHIC energy are, 17.3, 32.2 and 27.4 $fm/c$, respectively 
\cite{bassDumitru}. The hadronic rescattering cross sections 
for $\Omega$ and $\Xi$ particles and also their mean freeze-out time 
are much less compared to $\Lambda$ particles \cite{bleicher-bass}.  This shows that $\Omega$ comes out of the fireball
almost without interacting with the hadronic medium and hence the
enhancement is much less affected compared to $\Xi$ and $\Lambda$ particles. Additionally,
UrQMD calculations with a color flux tube break up mechanism having strong color fileds predict a dramatic 
enhancement in the multi-strange hadron production compared to $p+p$ interactions, $e.g.$ 
$\Lambda$'s are enhanced by a factor of 7, whereas $\Omega$'s are enhanced by a factor of 60 \cite{bleicher}. 

\begin{figure}
%Fig7
\includegraphics[width=86mm]{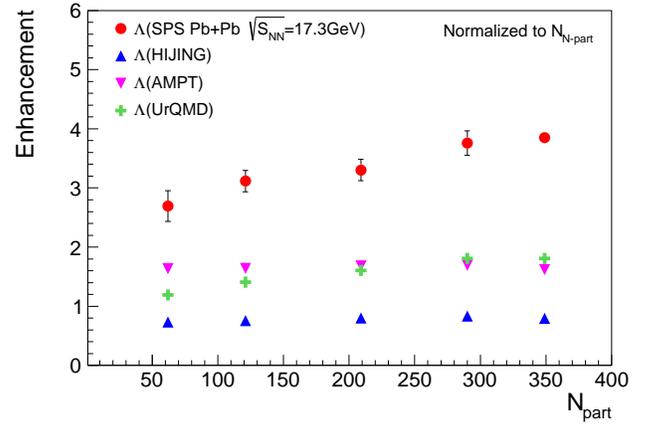}\\
\caption{\small (Color online) $N_{N-part}$-normalized
  enhancement of $\Lambda$ as a function of collision centrality for Pb+Pb collisions at $\sqrt{s_{NN}} = 17.3$
  GeV are compared with HIJING, AMPT and UrQMD models at mid-rapidity.}
\label{lambdaSimSps}
\end{figure}

\begin{figure}
%Fig8
\includegraphics[width=86mm]{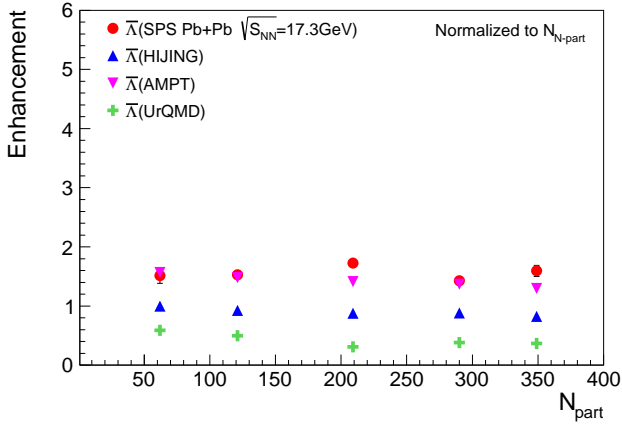}\\
\caption{\small (Color online) $N_{N-part}$-normalized
 enhancement of $\bar{\Lambda}$ as a function of collision centrality for Pb+Pb collisions at $\sqrt{s_{NN}} = 17.3$
 GeV are compared with HIJING, AMPT and UrQMD models at mid-rapidity.}
\label{aLambdaSimSps}
\end{figure}

The enhancement ratio of baryons and anti-baryons is usually affected by the net-baryon content
or baryon stopping at mid-rapidity.  This difference decreases with increase in collision energy.
For this reason, the enhancement of baryons and anti-baryons are different at SPS energies
which is not that much at RHIC energies.
In Figure \ref{lambdaSimSps} and Figure \ref{aLambdaSimSps}, $N_{N-part}$-normalized enhancement of
$\Lambda$ and $\bar{\Lambda}$-baryons for Pb+Pb collisions at $\sqrt{s_{NN}} = 17.3 $ GeV 
are plotted as a function of $N_{N-part}$, respectively. The collision data at
$\sqrt{s_{NN}} = 17.3 $ GeV are from SPS NA57 experiment \cite{spsStrange}
and are compared
with the corresponding estimates of HIJING, AMPT and UrQMD models, as shown in 
Figure \ref{lambdaSimSps} and Figure \ref{aLambdaSimSps}. The HIJING and AMPT data show
a centrality independent behavior, whereas UrQMD shows a weak centrality dependence. But the
collision data of $\Lambda$ show a remarkable centrality dependence, as shown in 
Figure \ref{lambdaSimSps}. At the same time the collision data, HIJING, AMPT and UrQMD
data of $\bar{\Lambda}$ are almost independent of centrality, which are shown in 
Figure \ref{aLambdaSimSps}. The difference in behavior of $\Lambda$ and $\bar{\Lambda}$
at $\sqrt{s_{NN}} = 17.3 $ GeV could be due to different
production mechanisms. This difference
 in the  production mechanism of $\bar{\Lambda}$ has been assigned to
 the following reasons. Firstly, the production threshold for anti-hyperons is
 larger than hyperons. Secondly, while going from higher to lower
 collision energy, baryon density of the system increases. This
 doesn't favor the production of particles without having a common
 valence quark with nucleons \cite{spsStrange}.
This difference in yields go away as one moves from SPS energy to top RHIC energy.

\begin{figure}
%Fig9
\includegraphics[width=86mm]{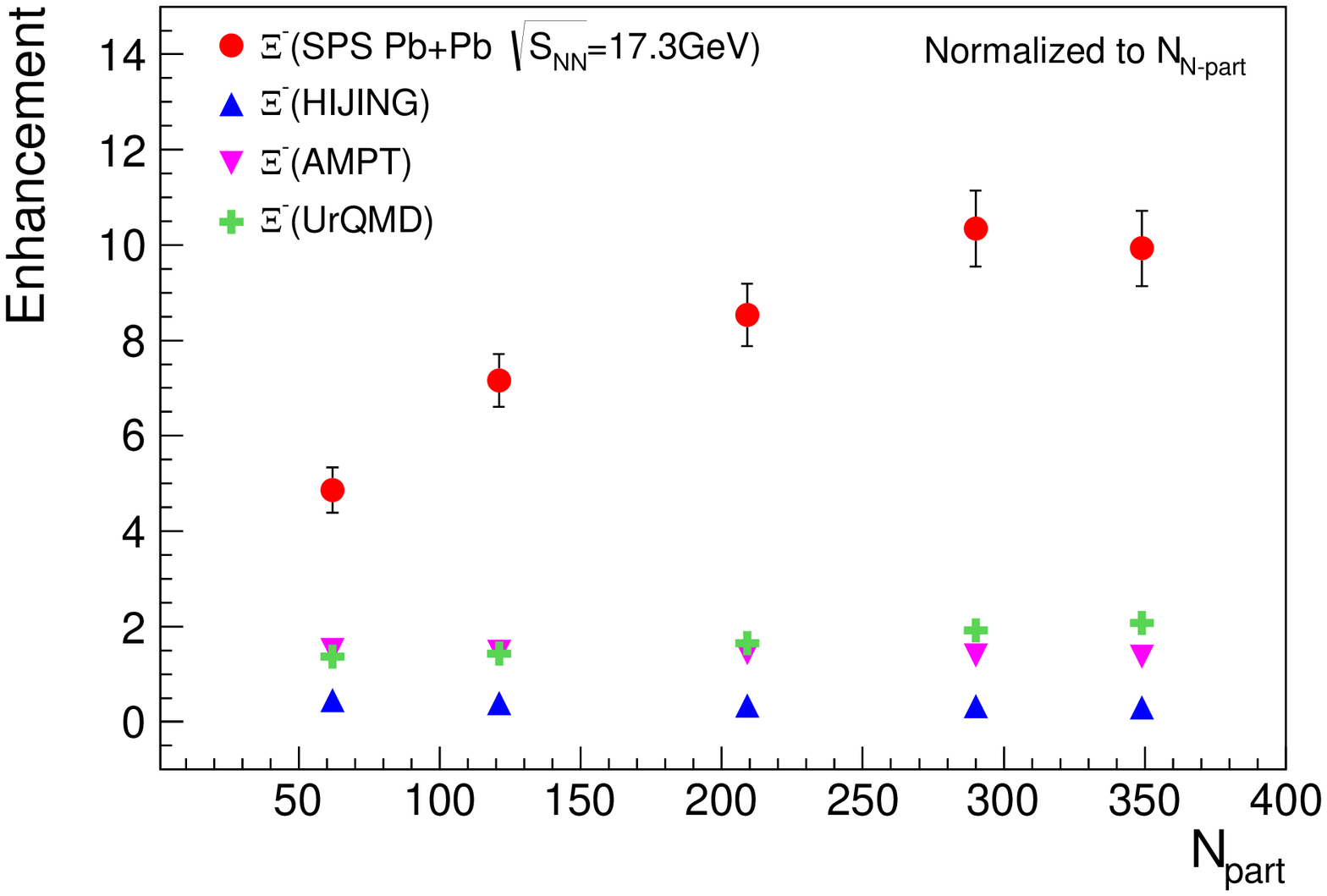}\\
\caption{\small (Color online) $N_{N-part}$-normalized
  enhancement of $\Xi^-$ as a function of collision centrality for Pb+Pb collisions at $\sqrt{s_{NN}} = 17.3$
  GeV are compared with HIJING, AMPT and UrQMD models at mid-rapidity.}
\label{xiSimSps}
\end{figure}

\begin{figure}
%Fig10
\includegraphics[width=86mm]{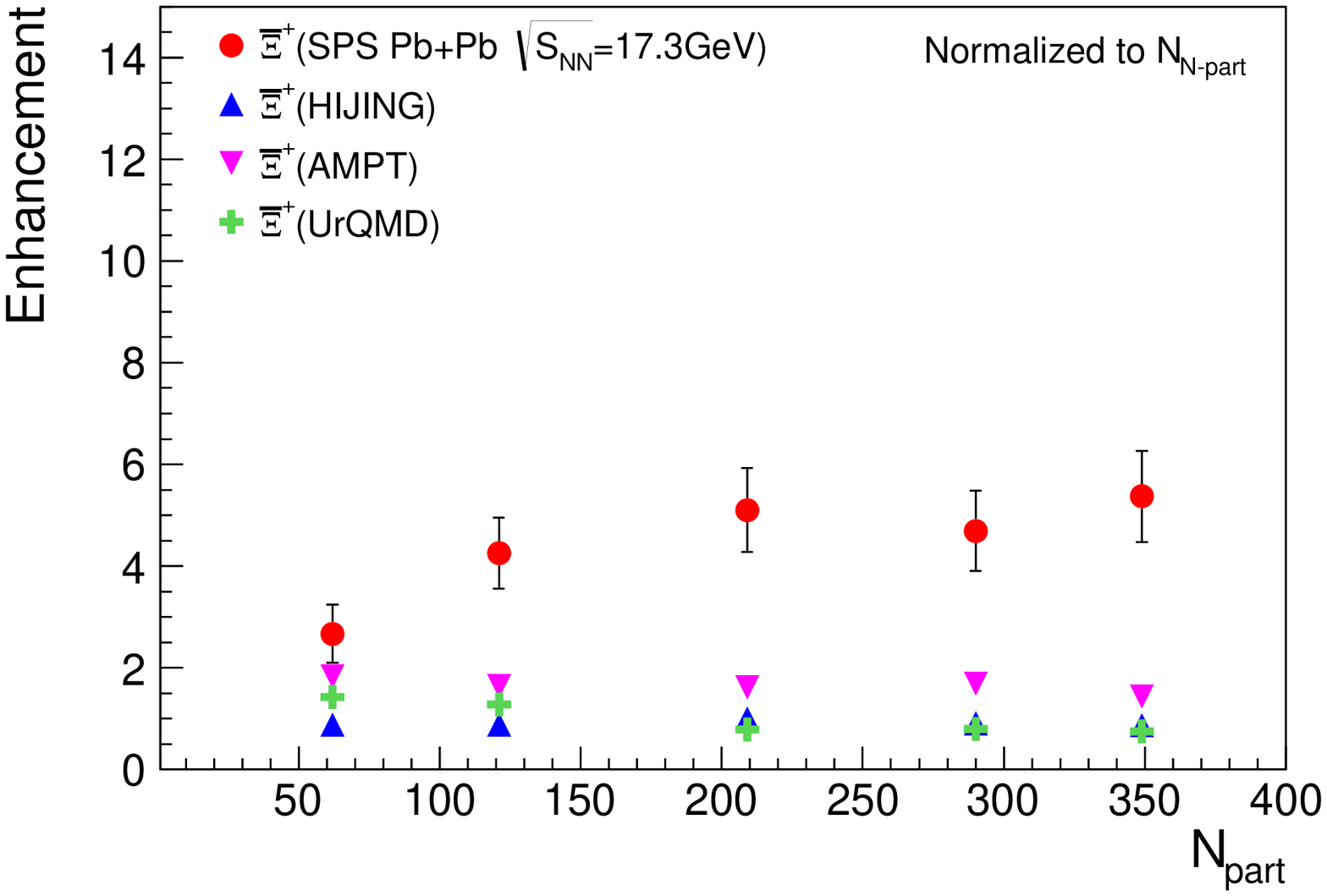}\\
\caption{\small (Color online) $N_{N-part}$-normalized
  enhancement of $\bar{\Xi}^+$ as a function of collision centrality for Pb+Pb collisions at $\sqrt{s_{NN}} = 17.3$
  GeV are compared with HIJING, AMPT and UrQMD models at mid-rapidity.}
\label{aXiSimSps}
\end{figure}

In Figure \ref{xiSimSps} and Figure \ref{aXiSimSps}, $N_{N-part}$-normalized enhancement of
$\Xi^-$ and $\bar{\Xi}^+$-baryons at $\sqrt{s_{NN}} = 17.3 $ GeV for Pb+Pb collisions for
SPS data have been shown as a function of $N_{N-part}$. These data
points are from NA57 experiment  and are compared
with the corresponding estimates of HIJING, AMPT and UrQMD models \cite{spsStrange}. 
At SPS energies, both $\Xi^-$ and $\bar{\Xi}^+$ show strangeness enhancement which 
rises with centrality, unlike the models under discussion. However, all the models with 
partonic and hadronic degrees of freedom show no enhancement of 
$\Xi^-$ and $\bar{\Xi}^+$-baryons and are independent of collision centrality.

\begin{figure}
%Fig11
\includegraphics[width=86mm]{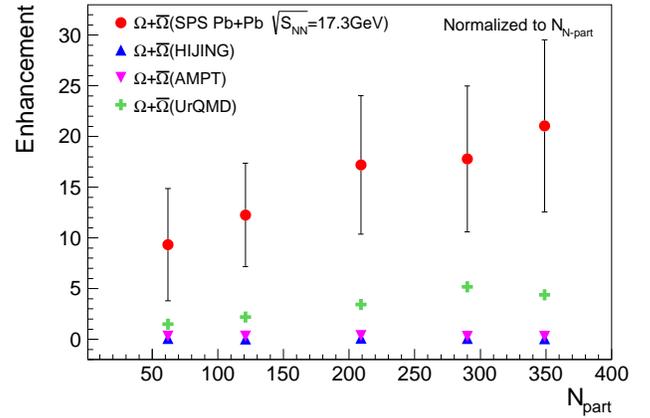}\\
\caption{\small (Color online) $N_{N-part}$-normalized
  enhancement of $\Omega+\bar{\Omega}$ as a function of centrality
for Pb+Pb collisions at $\sqrt{s_{NN}} = 17.3$  GeV are compared with HIJING, AMPT and UrQMD models at mid-rapidity.}
\label{omegaSimSps}
\end{figure}

In Figure \ref{omegaSimSps},  $N_{N-part}$-normalized enhancement of
$\Omega + \bar{\Omega}$-baryons at $\sqrt{s_{NN}} = 17.3 $ GeV for
Pb+Pb collisions have been shown as a function of $N_{N-part}$. These
data points are from NA57 experiment and are compared
with the corresponding estimates of HIJING, AMPT and UrQMD models \cite{spsStrange}. 
 A similar observation has been made for $\Omega + \bar{\Omega}$-baryons 
 like the $\Xi$-baryons. To understand different enhancement profiles of multi-strange 
 baryons at SPS energies, it is important to study their production dynamics. However,
 it could be observed that the shape of the enhancements for (anti) baryons are similar,
 which goes inline with the predictions of a grand canonical ensemble approach (GC) 
 \cite{jc}.

\begin{figure}
%Fig12
\includegraphics[width=86mm]{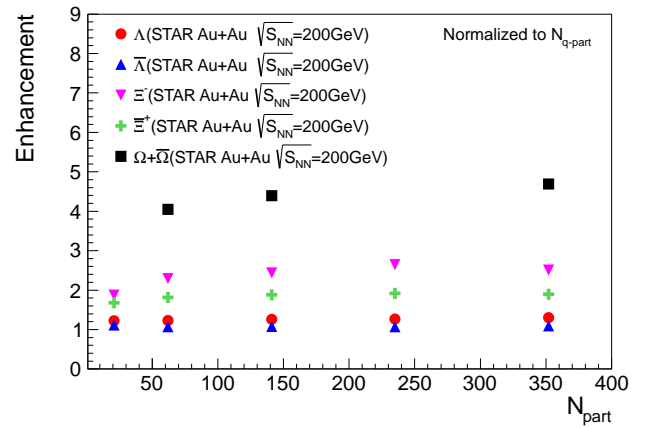}\\
 \caption{\small (Color online) Mid-rapidity $N_{q-part}$-normalized
   enhancement of multi-strange baryons as a function of centrality
   for Au+Au collisions at $\sqrt{s_{NN}} = 200$ GeV at RHIC.}
 \label{QenhanceRHIC}
\end{figure}

\begin{figure}
%Fig13
\includegraphics[width=86mm]{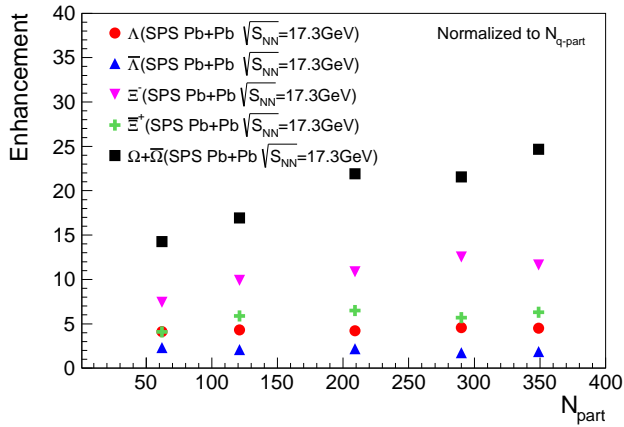}\\
 \caption{\small (Color online) Mid-rapidity $N_{q-part}$-normalized
   enhancement of multi-strange baryons as a function of centrality
   for Pb+Pb collisions at $\sqrt{s_{NN}} = 17.3$ GeV at SPS.}
 \label{QenhanceSPS}
\end{figure}

In Figure \ref{QenhanceRHIC},  $N_{q-part}$-normalized enhancement for multi-strange
baryons for Au+Au collisions at $\sqrt{s_{NN}} = 200 $ GeV have been 
shown as a function of collision centrality. These data points are
from the STAR experiment at RHIC \cite{strangeStar}. Here we observe that when the enhancement
is normalized to the number of constituent 
quarks, it turns out to be a centrality independent scaling behavior. This enhancement, however, 
depends on the particle mass and shows an increase with higher mass number. The former 
observation is very interesting in view of partonic degrees of freedom playing a crucial 
role in particle production especially for the multi-strange particles, the enhancement 
of which has been conjectured to be a signal of formation 
of a partonic phase. In other words, at top RHIC energy the collision could be 
described at a partonic level interactions. The number of quark participant 
scaling (NQ-scaling) works fine at intermediate 
$p_T$ for the elliptic flow of multi-strange particles, showing partonic collectivity 
at RHIC \cite{multiNQ}. This supports present observation of centrality independent scaling behavior of 
constituent quarks normalized multi-strange baryon enhancement.
However, the top SPS energy shows a weak quark participant scaling
towards higher strangeness content of the particles, which is shown in 
Figure \ref{QenhanceSPS}. Hence, it could be inferred  here that an
onset of deconfinement phase transition might have taken place already at top SPS energies leading to a mixed phase
of partons and hadrons. This goes inline with other experimental 
findings related to the 
onset of deconfinement or saturation of different observables starting from SPS 
energies \cite{onset,saturation}.

In Figure \ref{QenhanceSPS},  $N_{q-part}$-normalized enhancement for multi-strange 
baryons at  $\sqrt{s_{NN}} = 17.3 $ GeV for Pb+Pb collisions have been
shown as a function of collision centrality. These data points are
from NA57 experiment \cite{spsStrange}. Here, we observe that 
for lower strangeness content particles, the quark participant normalized multi-strange 
enhancement turns out to be a centrality independent scaling-behavior. However, this is not 
true specifically for the $\Omega$ baryons, for which we observe a linear rise. This 
makes a difference between SPS and RHIC energies so far the constituent quark scaling 
of multi-strange baryons goes.

\begin{figure}
%Fig14
\includegraphics[width=86mm]{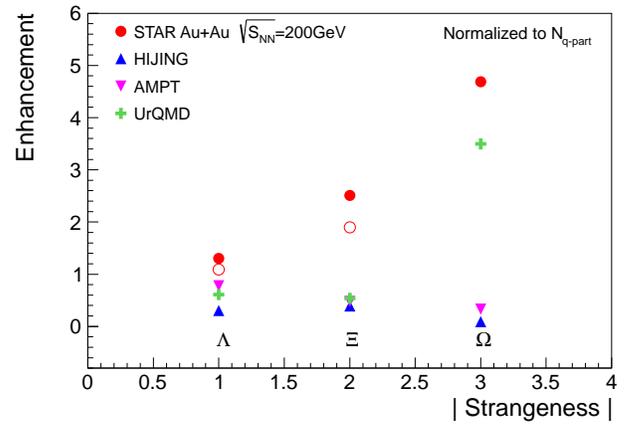}\\
\caption{\small (Color online) Mid-rapidity $N_{q-part}$-normalized enhancement of
  multi-strange baryons as a function of strangeness content for Au+Au
  collisions at $\sqrt{s_{NN}} = 200$ GeV at RHIC. The filled and open
  circles are for the particles and anti-particles, respectively.}
\label{QenhanceS-rhic}
\end{figure}

\begin{figure}
%Fig15
\includegraphics[width=86mm]{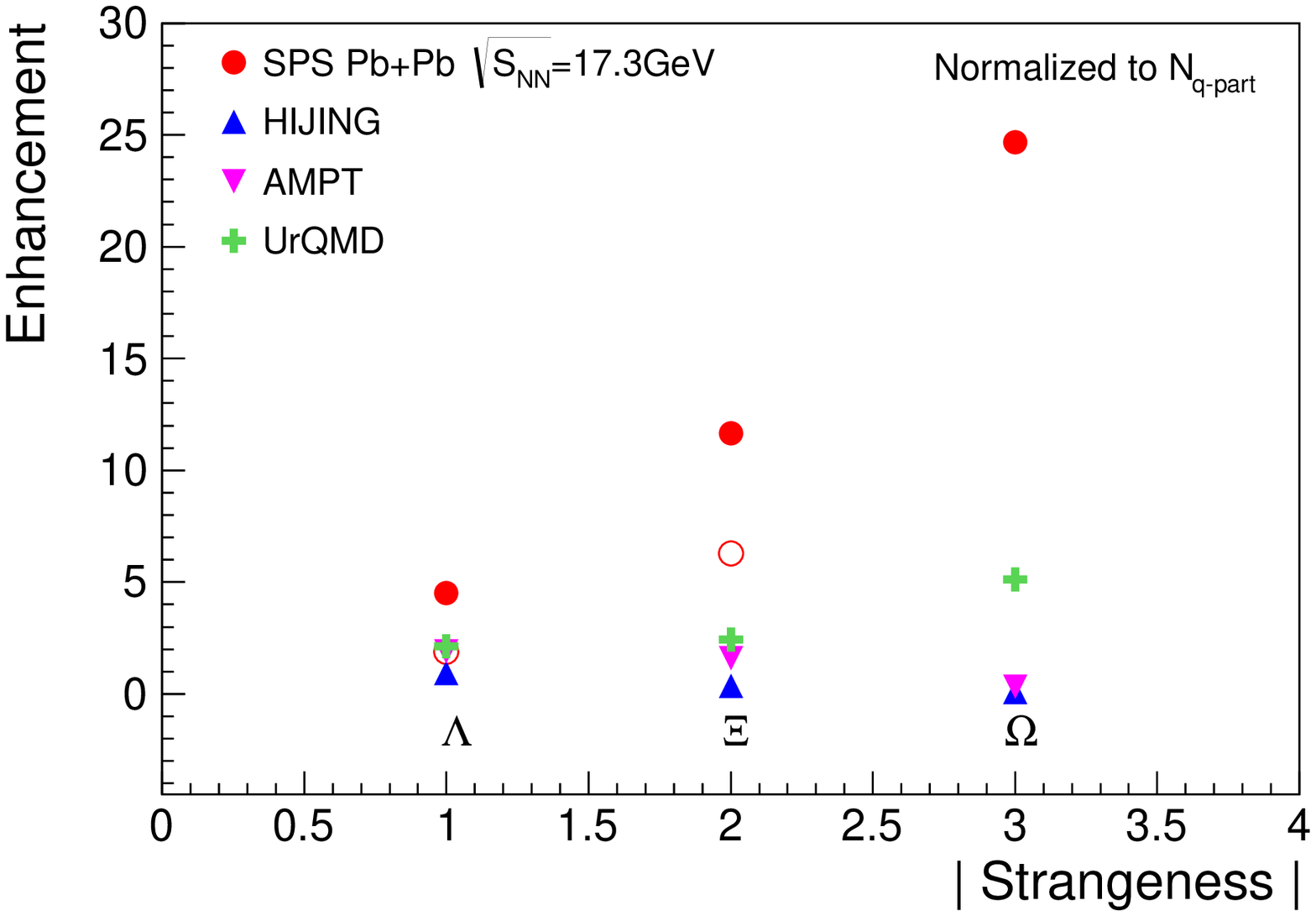}\\
\caption{\small (Color online) Mid-rapidity $N_{q-part}$-normalized enhancement of
  multi-strange baryons as a function of strangeness content for Pb+Pb
  collisions at $\sqrt{s_{NN}} = 17.3$ GeV at SPS. The filled and open circles are for the particles and anti-particles, respectively.}
\label{QenhanceS-sps}
\end{figure}
\begin{figure}
%Fig16
\includegraphics[width=86mm]{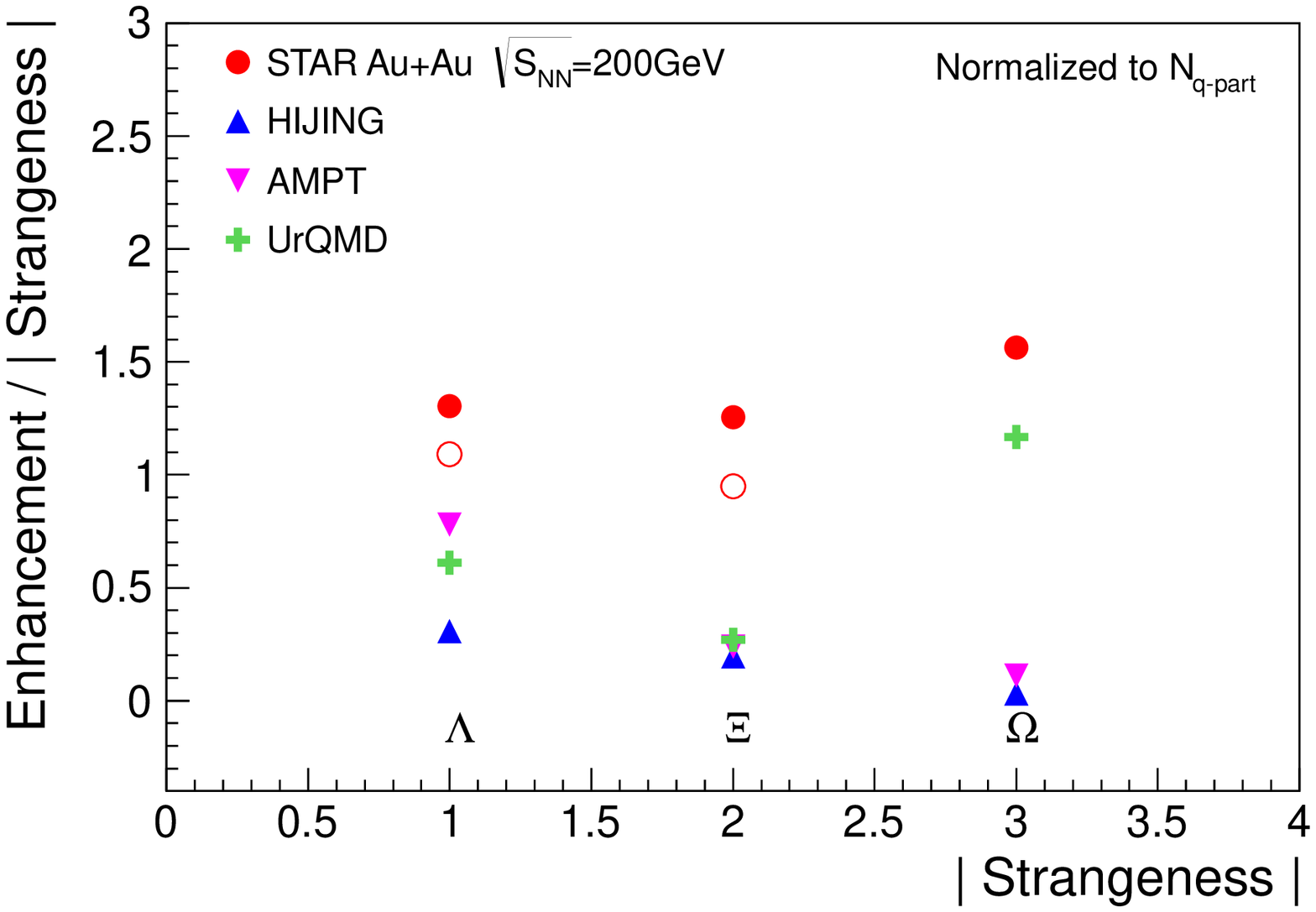}\\
\caption{\small (Color online) Mid-rapidity $N_{q-part}$ and
  strangeness content normalized enhancement of
  multi-strange baryons as a function of strangeness content for Au+Au
  collisions at $\sqrt{s_{NN}} = 200$ GeV at RHIC. The filled and open circles are for the particles and anti-particles, respectively.}
\label{QenhanceS-rhic-S}
\end{figure}

\begin{figure}
%Fig17
\includegraphics[width=86mm]{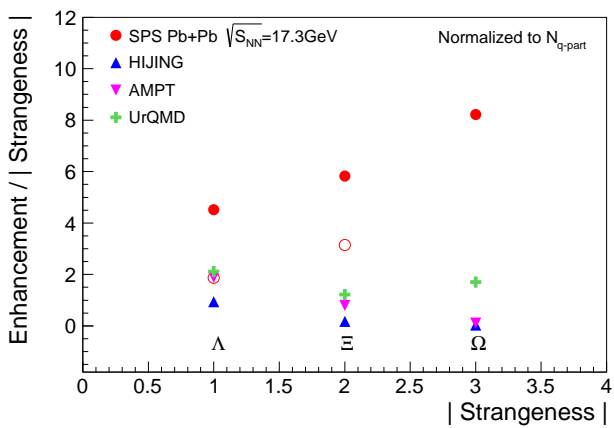}\\
\caption{\small (Color online) Mid-rapidity $N_{q-part}$ and
  strangeness content normalized enhancement of
  multi-strange baryons as a function of strangeness content for Pb+Pb
  collisions at $\sqrt{s_{NN}} = 17.3$ GeV at SPS. The filled and open circles
   are for the particles and anti-particles, respectively.}
\label{QenhanceS-sps-S}
\end{figure}

 Statistical model estimation of strangeness enhancement in a broad range of energies spanning from 
 $\sqrt{s_{NN}} =8.73-130$ GeV suggests that (i) the absolute value of the enhancement decreases with increasing collision energy and increases with strangeness content, (ii) the enhancement pattern as a function of centrality seems to be preserved at all energies \cite{jc}. The fact that the chemical freeze-out temperature is almost constant at higher energies and there is little effect of baryon chemical potential on the strength and pattern of strangeness enhancement, going from 130 to 200 GeV collision energy won't change these observations. At lower collision energies, the initial conditions don't favor a deconfinement transition. Hence, as discussed in Ref. \cite{jc}, strangeness enhancement and the pattern of enhancement are not a unique signal of deconfinement, which is a consequence of canonical suppression of strangeness in $p+p$ collisions. However, a quark participant scaling of strangeness enhancement could be considered as a signal of deconfinement transition and formation of a QGP phase, which is seen for the top RHIC energy.  A weak quark participant scaling of strangeness enhancement seen at the top SPS energy in turn indicates a co-existence of partonic and hadronic phases.

The absolute values 
of the enhancement factor (after normalizing to the number of constituent quarks) as a 
function of strangeness content of different multi-strange baryons at top RHIC and SPS 
energies are shown in Figures \ref{QenhanceS-rhic} and \ref{QenhanceS-sps}, respectively. 
None of the models used here predicts any enhancement for $\Omega$s except UrQMD. 
The reasons for this are discussed earlier in this paper.
It is evident from these plots that the absolute enhancement number
decreases with collision energy. This could be understood in a grand canonical ensemble
approach, which predicts a significant decrease in (anti) baryon enhancements with collision
energy \cite{jc,strangeStar}. We do observe a monotonic rise in the number of constituent quarks 
normalized strangeness enhancement as a function of strangeness content for both the energies 
({\it i.e.}  $E(\Omega) > E(\Xi) > E(\Lambda)$).  This could be understood as follows. 
The strangeness enhancement in A+A collisions is always measured relative to $p+p$ or $p+A$ collisions 
after appropriately dividing by the number of participants.  In $p+p$ collisions, even at 
higher energies, the available thermal or energy phase space is very less compared to A+A collisions. 
The absolute yields of strange particles in $p+p$ collisions go down with increase in strangeness content because
of canonical suppression. Compared to A+A collisions, the gluon density in $p+p$ collisions is very less, 
which contributes to strangeness production ($gg \rightarrow s\bar{s}$) more abundantly. 
In A+A collisions, there is enough phase space available for the formation of multi-strange 
particles.  When we look at strangeness enhancement, particles with higher strangeness 
content are enhanced more, compared to the enhancement of particles with lower strangeness 
content \cite{jc}. This also talks about the effect of the medium produced in A+A collisions, where quark 
coalescence and gluon fusion play important roles in multi-strange particle production.
However, when the enhancement factor is further normalized with respect to  the strangeness 
content and is plotted as a function of the strangeness content,  a monotonic
rise is observed for top SPS energy. Whereas it remains almost flat for the top RHIC energy. 
This means at top most RHIC energy both constituent quark and strangeness scaling of 
multi-strange baryons have been observed. It can be observed from 
Figures \ref{QenhanceRHIC} and \ref{QenhanceSPS} that the $N_{q-part}$ scaling is more prominent 
at RHIC compared to SPS. In view of the fact that onset of deconfinement starts at SPS energies and 
there are several observations of a mixed phase of partons and hadrons at SPS energies \cite{onset,onsetDe},
 a weak $N_{q-part}$ scaling and strangeness scaling at these energies are expected. However, RHIC 
 shows a strangeness scaling along with $N_{q-part}$ scaling. This gives an evidence of a purely partonic phase.
 These are shown in Figures \ref{QenhanceS-rhic-S} and \ref{QenhanceS-sps-S} . 
 We expect that these scaling laws will hold good at LHC energies. 
 
%-------------------->
\section{SUMMARY AND CONCLUSIONS}

In this work, we have calculated 
the number of nucleon and quark participants in a proton-proton, proton-nucleus and  
nucleus-nucleus collisions in the framework of a nuclear overlap model. 
The data for the enhancement of multi-strange baryons (in A+A collisions compared to $p+p$ or $p+Be$)
at top SPS and RHIC energies have been compared with HIJING, UrQMD and AMPT models.  
We observe the $N_{N-part}$-normalized yields of multi-strange
particles show a monotonic increase with centrality. This turns out to be a centrality independent scaling 
behavior when normalized to number of constituent quarks participating in the 
collision. This geometrical scaling indicates the partonic degrees of 
freedom playing important role in the production of multi-strange 
particles. Furthermore, we see the constituent quark normalized yield when 
further divided by the strangeness content, it shows a strangeness 
independent scaling behavior at the top RHIC energy, whereas this scaling is not
 seen at the top SPS energy. This goes in line with other observations at RHIC towards 
 the formation of a partonic phase and a mixed phase of partons and hadrons at 
 SPS energies. These features are reproduced neither by explicit hadronic kinetic models
like  UrQMD, HIJING nor by AMPT model which treats the partonic phase on the basis of 
pQCD with massless partons and a noninteracting equation-of-state. A quark participant scaling of strangeness enhancement, in view of the above observations, seems to signal a deconfined state of quark-gluon plasma.

\begin{acknowledgments}
Authors would like to thank Dr. Dariusz Miskowiec, GSI, Germany  for
fruitful discussions on the nuclear verlap model and Prof. R. Varma, IIT Bombay for useful discussions. 
\end{acknowledgments}

\vskip-0.75cm
% Create the reference section using BibTeX:

\end{document}